\documentclass[
reprint,superscriptaddress, amsmath,amssymb, aps,
]{revtex4-1}

\usepackage{graphicx}
\usepackage{dcolumn}
\usepackage{bm}
\usepackage{ulem}
\usepackage[usenames, dvipsnames]{xcolor}

\begin{document}

\title{Electrostatic control of quasiparticle poisoning \\ in a hybrid semiconductor-superconductor island}

\author{H.~Q.~Nguyen}
 
\affiliation{Center for Quantum Devices, Niels Bohr Institute, University of Copenhagen, 2100 Copenhagen, Denmark}
 \affiliation{Nano and Energy Center, Hanoi University of Science, VNU, 120401 Hanoi, Vietnam}
 
 \author{D.~Sabonis}

\affiliation{Center for Quantum Devices, Niels Bohr Institute, University of Copenhagen, 2100 Copenhagen, Denmark}

\author{D.~Razmadze}
\affiliation{Center for Quantum Devices, Niels Bohr Institute, University of Copenhagen, 2100 Copenhagen, Denmark}
 
\author{E.~T.~Mannila}
\affiliation{QTF Centre of Excellence, Department of Applied Physics, Aalto University, FI-00076 Aalto, Finland}

\author{V.~F.~Maisi}
\affiliation{QTF Centre of Excellence, Department of Applied Physics, Aalto University, FI-00076 Aalto, Finland}
\affiliation{Division of Solid State Physics and NanoLund, Lund University, 22100 Lund, Sweden}
 
\author{D.~M.~T.~van~Zanten}
\affiliation{Center for Quantum Devices, Niels Bohr Institute, University of Copenhagen, 2100 Copenhagen, Denmark}

\author{E.~C.~T.~O'Farrell}
\affiliation{Center for Quantum Devices, Niels Bohr Institute, University of Copenhagen, 2100 Copenhagen, Denmark}

\author{P.~Krogstrup}
\affiliation{Center for Quantum Devices, Niels Bohr Institute, University of Copenhagen, 2100 Copenhagen, Denmark}

\author{F.~Kuemmeth}
\affiliation{Center for Quantum Devices, Niels Bohr Institute, University of Copenhagen, 2100 Copenhagen, Denmark}
 
\author{J.~P.~Pekola}
\affiliation{QTF Centre of Excellence, Department of Applied Physics, Aalto University, FI-00076 Aalto, Finland}

\author{C.~M.~Marcus}
\affiliation{Center for Quantum Devices, Niels Bohr Institute, University of Copenhagen, 2100 Copenhagen, Denmark} 

\begin{abstract}
The performance of superconducting devices is often degraded by the uncontrolled appearance and disappearance of quasiparticles, a process known as poisoning. We demonstrate electrostatic control of quasiparticle poisoning in the form of single-charge tunneling across a fixed barrier onto a Coulomb island in an InAs/Al hybrid nanowire.  High-bandwidth charge sensing was used to monitor charge occupancy of the island across Coulomb blockade peaks, where tunneling rates were maximal, and Coulomb valleys, where tunneling was absent. Electrostatic gates changed on-peak tunneling rates by two orders of magnitude for a barrier with fixed normal-state resistance, which we attribute to gate dependence of the size and softness of the induced superconducting gap on the island, corroborated by separate density-of-states measurements. Temperature and magnetic field dependence of tunneling rates are also investigated. 
\end{abstract}

\pacs{xyz}
\maketitle
Recent advances in hybrid semiconductor-superconductor materials \cite{Lutchyn} has led to new modalities of control of superconducting devices from multiplexers to detectors to qubits. For instance, in hybrid nanowires (NWs), the combination of superconductivity, spin-orbit interactions, and Zeeman coupling can give rise to Majorana zero modes \cite{OregPRL10, LutchynPRL10,MourikScience,AlbrechtNature}, expected to exhibit non-abelian braiding statistics potentially useful for error-protected quantum computing \cite{Alicea2}. For this and other applications \cite{deLangePRL, LarsenPRL, CasparisNNano, HaysScience} it is vital to engineer long parity lifetime in these new systems \cite{HigginbothamParity}. 

\begin{figure}[t]
\begin{center}
\includegraphics[width=3in,keepaspectratio]{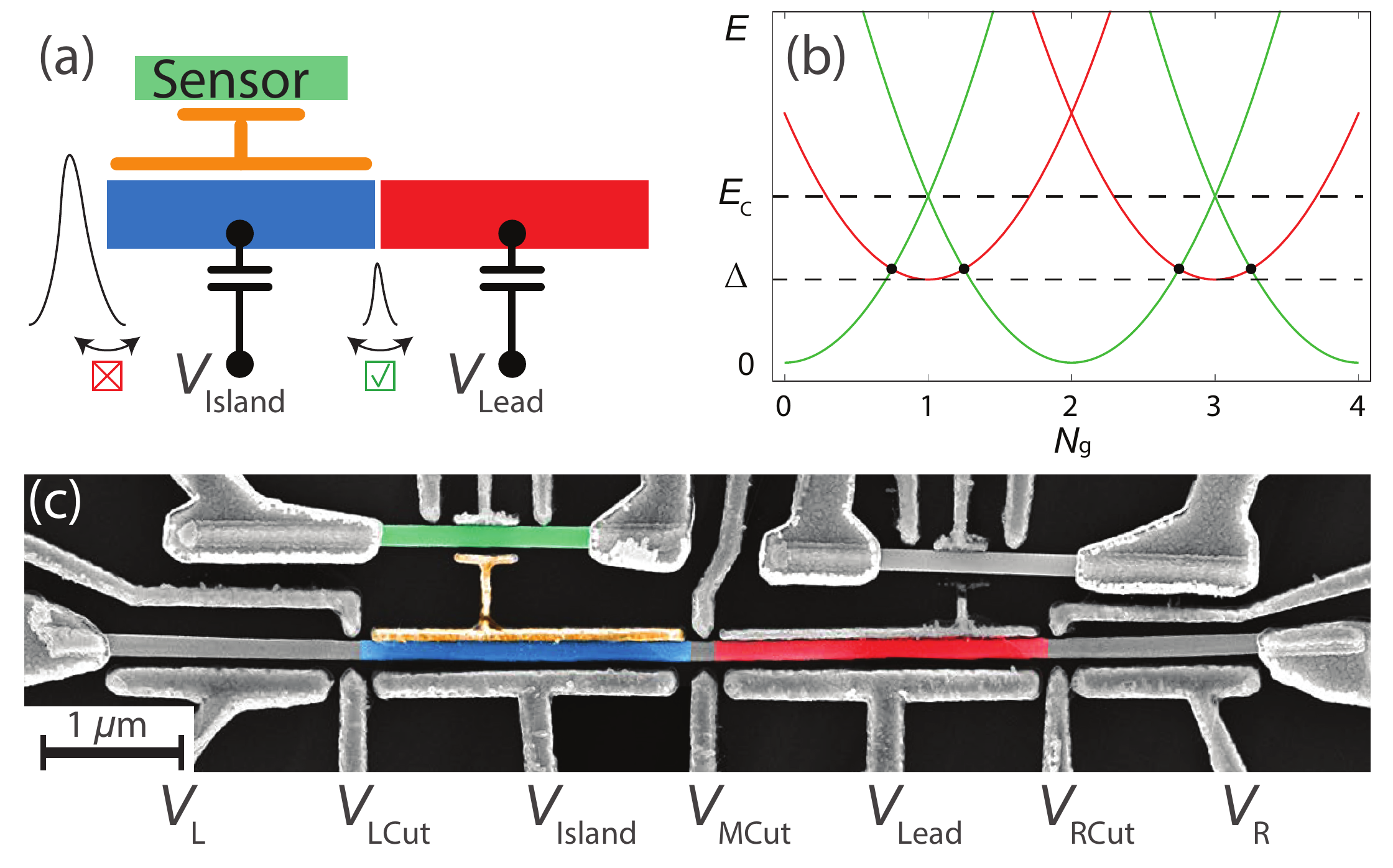}
\caption{(a) Schematic of the measurement configuration: the superconducting island (blue) is capacitively coupled to an RF charge sensor (green) using a floating metallic coupler (yellow). High-bandwidth readout allows detection of the charge state of the island with time resolution of a few microseconds. (b) Electrostatic energy diagram of the superconducting island in the even-odd regime when the charging energy $E_\text{C}$ is larger than the induced superconducting gap $\Delta$. (c) False-color  electron micrograph of the nanowire with colored segments corresponding to the schematic in (a). Long plunger gates control the electron density of the corresponding nanowire segment while the short cutter gates control tunnel barriers. 
Quasiparticle tunneling is controlled by $V_{\text{Lead}}$.} 
\label{Fig1}
\end{center}
\end{figure}

A superconducting island coupled to electronic reservoirs via tunneling barriers has a ground state with all electrons paired whenever the superconducting gap $\Delta$ exceeds the charging energy $E_\text{C}$. On the other hand, if $E_\text{C}> \Delta$, charge states involving an unpaired electron can become energetically favorable, and ground states show alternating even-odd charge occupation as a function of gate-induced charge $N_g$ [See Fig.~\ref{Fig1}(b)]. At elevated temperatures or out of equilibrium, unpaired quasiparticles (QPs) generated within the device or entering via tunneling restore 1$e$ periodicity via a process termed QP poisoning.

Experiments have previously shown that intentionally engineering the superconducting gaps of the island $\Delta_{\rm{Island}}$, and lead $\Delta_{\rm{Lead}}$, to be unequal can strongly influence the tunneling rates of the island \cite{PekolaAPL09,AumentadoPRL04}. In particular, for $\Delta_{\rm{Lead}} < \Delta_{\rm{Island}}$, where, on average, QPs should be repelled from the island, it was found experimentally that low-temperature Coulomb blockade was 2$e$ periodic. When $\Delta_{\rm{Island}} < \Delta_{\rm{Lead}}$, where QPs should be, on average, attracted to the island, $1e$ periodicity was observed, indicating rapid poisoning of the island from the lead.

In this Letter, we investigate on-resonance tunneling of $1e$ charge onto and off of a tunnel-coupled Coulomb island (QP poisoning) in an epitaxial InAs/Al NW device with an integrated charge sensor and a lead made from the same NW, with separate gates controlling the potential and density on the island and the density in the semiconductor part of the lead. The island has $E_\text{C}>\Delta$ \cite{ChangNNano, HigginbothamParity, AlbrechtNature}. We find that deep in the Coulomb blockade valley, the charge configuration was stable and no tunneling was observed. Close to a charge transition of the island, where protection by $E_\text{C}$ is lifted,  QP tunneling was observed in real time. We found that the tunneling rate at the charge transition was controllable over two orders of magnitude by gating the lead.  Correlating this behavior with bias spectroscopy suggests that it is the influence of gate voltages on the induced gap on the island and leads that is responsible for the gate dependent tunneling, comparable to Refs.~\cite{PekolaAPL09,AumentadoPRL04}. Increased tunneling with magnetic field and temperature was also investigated.  

Tunneling of QPs on $\mu$s to ms timescales have previously been detected in real time using fast radio frequency (RF) reflectometry \cite{FergusonPRL06, ClarkPRB08, Ferguson17, DelsingPRB08, BylanderNature05, MaisiPRL14,GerboldPRB,RazmadzePRN}. QP poisoning rates of superconducting islands have previously been estimated based on statistics of switching current while changing the current ramping rate \cite{AumentadoPRB06, KouwenhovenNPhys15, Veenarxiv18}. Here we implement a more direct method, by directly reading the island charge using a high-bandwidth integrated charge sensor \cite{SchoelkopfScience98, RazmadzePRN}.

Figure \ref{Fig1}(a) shows a schematic of the measurement configuration. The Coulomb blockaded superconducting island (blue) is capacitively coupled to an RF charge sensor (green) using a floating metallic coupler (yellow). The energy diagram of the superconducting island is shown in Fig.~\ref{Fig1}(b). The even-odd regime is characterized by alternating spacing between the charge degeneracies when $E_\text{C}>\Delta$.  The superconducting island was tunnel coupled to a superconducting lead (red), fabricated on the same NW, ohmically connected to a normal-metal reservoir. Figure \ref{Fig1}(c) shows a false-color micrograph of the device. The 100 nm diameter NW is grown using the vapor-liquid-solid technique in a molecular beam epitaxy system with the InAs [111] substrate crystal orientation. Following the NW growth, Al is deposited epitaxially \textit{in situ} on three facets of the NW with an average thickness of 10 nm \cite{Krogstrup}. The NW is then manually positioned on a chip with few-$\mu$m precision. Using electron beam lithography and Transene D wet etch, the Al shell was removed from the nanowire near narrow gates denoted LCut, MCut, and RCut (cutters). Extended gates denoted L, Island, Lead, and R (plungers) tune the density in the corresponding segment of the NW \cite{Sole2018, deMoor2018}. In the measurement configuration, the voltage on gate LCut was set strongly negative such that tunneling to the island was controlled entirely by $V_{\rm MCut}$. Charge detection was performed using the capacitively coupled charge sensor with a 20~$\mu$s integration time \cite{RazmadzePRN}. 
During the counting experiment, the bias voltage across the superconducting island was set to zero, $V_\text{SD}=0$, thereby grounding the leads. 
The measurements were performed in a dilution refrigerator with a base temperature of 20 mK and a 1-1-6 T vector magnet.

\begin{figure}[t]
\begin{center}
\includegraphics[width=3.25in,keepaspectratio]{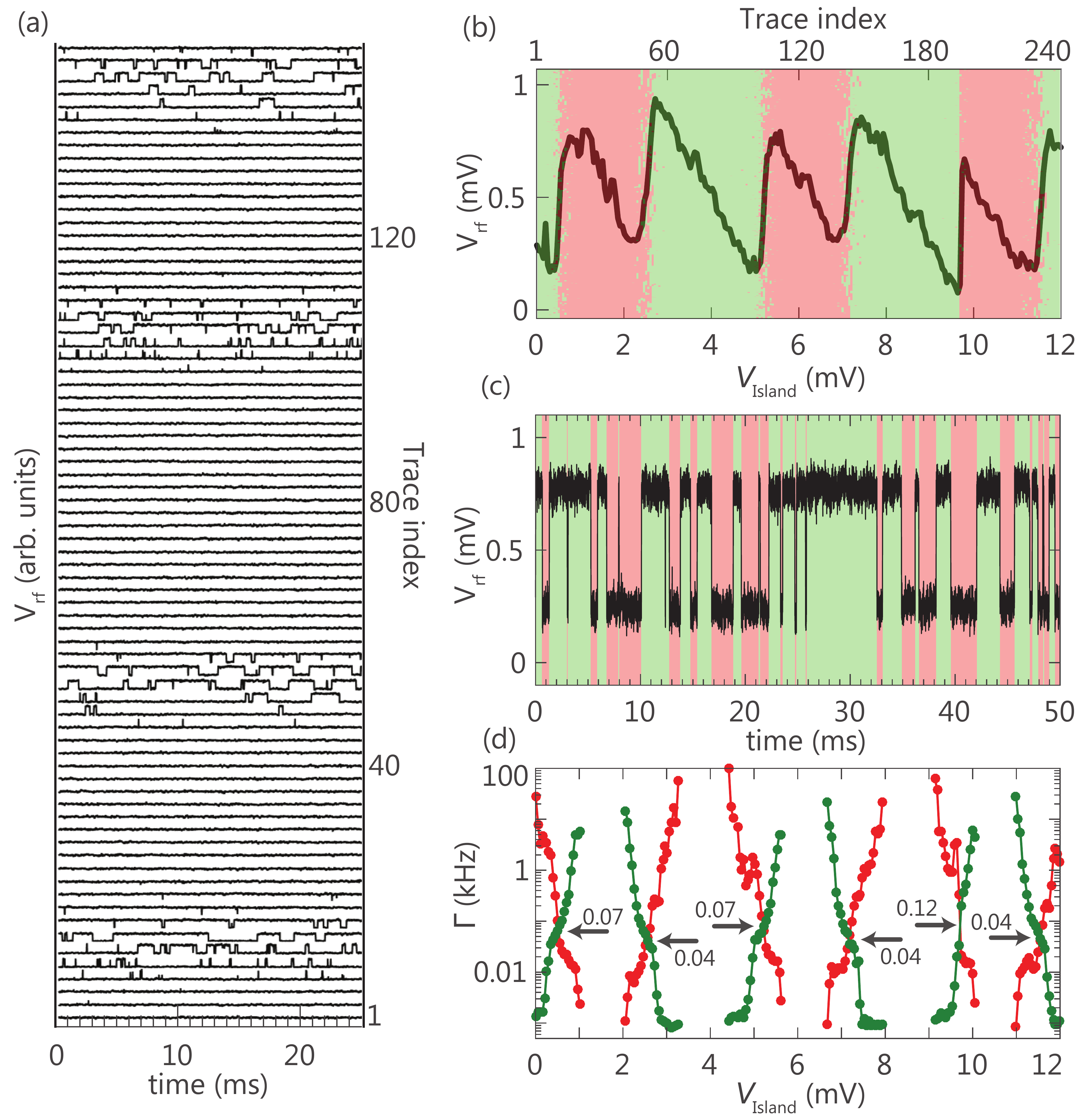}
\caption{(a) Time traces (rows offset) showing $1e$ tunneling of the island (blue segment in Fig.~\ref{Fig1}) as the plunger voltage $V_{\rm{Island}}$ was swept over several Coulomb valleys. Near charge degeneracies, individual switching events are visible in the demodulated RF signal $V_{\rm rf}$, while island charge is stable within Coulomb valleys. (b) Time-averaged charge states with color map indicating alternation of average charge. (c) A zoom-in time trace close to charge degeneracy showing single-electron tunneling in real time. The background colors show the digitized data binned to two levels using a thresholding algorithm and corresponding to one excess electron on or off the island. (d) Tunneling rates $\Gamma_{\text{e} \rightarrow \text{o}}$ (green) and $\Gamma_{\text{o} \rightarrow \text{e}}$ (red) as a function of island plunger voltage $V_{\rm{Island}}$. The gate dependence of the crossing point of the two rates show an even-odd effect.  Average equilibrium (on resonance) tunneling rate $\Gamma_{\rm{eq}}$ is found by averaging rates at several crossing points. All data at zero magnetic field.}
\label{Fig2}
\end{center}
\end{figure}

For all measurements, $V_L=V_R=0\,$V, and $V_{\rm{RCut}}$ was set positive to fully open the RCut junction. For charge-counting measurements, a single-lead Coulomb island was formed by setting $V_{\rm{LCut}}$ strongly negative, disconnecting the left side of the island, while $V_{\rm{MCut}}$ was set so that the normal-state conductance of MCut was $\sim$0.35~$e^2/h$, checked via transport with LCut fully open. Typical charge sensing data for $V_{\rm{Lead}}$ = 0 V are shown in Fig.~\ref{Fig2}(a), with each trace shifted for clarity. For each time trace, the demodulated reflectometry voltage $V_{\rm rf}$ was sampled at a fixed plunger voltage $V_{\rm{Island}}$, then $V_{\rm{Island}}$ was stepped to the next value. 
Near charge degeneracies rapid tunnelings were observed, while away from transitions the switching vanished, reflecting stable charge configurations in Coulomb valleys.  Averaging each time trace yielded a single average charge-sensor signal, which is plotted as a function of $V_{\rm{Island}}$ in Fig.~\ref{Fig2}(b). The moderate amount of even-odd spacing of the transitions suggest a charging energy $E_\text{C} \sim 500$ $\mu$eV, roughly twice the induced superconducting gap $\Delta \sim 250$ $\mu$eV \cite{EilesPRL93}.

Figure~\ref{Fig2}(c) shows a time trace acquired with $V_{\rm{Island}}$ fixed near a charge degeneracy. The high signal-to-noise ratio of the sensor signal (SNR~$>$~3 \cite{SNR}) allowed the use of simple thresholding to determine transitions between odd and even occupations, color coded in  Fig.~\ref{Fig2}(c), rather than more sophisticated thresholding techniques \cite{Ferguson17, PranceNano}. Specifically, tunneling rates were determined from time traces like this by dividing the number of transitions out of a charge state, even or odd, $N_{\text{e(o)} \rightarrow \text{o(e)}}$, by the total time $\sum t_{\text{e(o)}}$ spent in that state within the time trace, $\Gamma_{\text{e(o)} \rightarrow \text{o(e)}}=N_{\text{e(o)} \rightarrow \text{o(e)}}/\sum t_{\text{e(o)}}$ \cite{MaisiPRL11}.

Resulting tunneling rates $\Gamma_{\text{e} \rightarrow \text{o}}$ (green) and $\Gamma_{\text{o} \rightarrow \text{e}}$ (red) are shown in Fig.~\ref{Fig2}(d). The two rates cross at each charge degeneracy, identifying both the value of $V_{\text{Island}}$ where even and odd occupancies are equally likely, and the tunneling rate ($1e$ charge transition rate) at that transition, indicated by black arrows in Fig.~\ref{Fig2}(d). 

\begin{figure}[t]
\begin{center}
\includegraphics[width=3in,keepaspectratio]{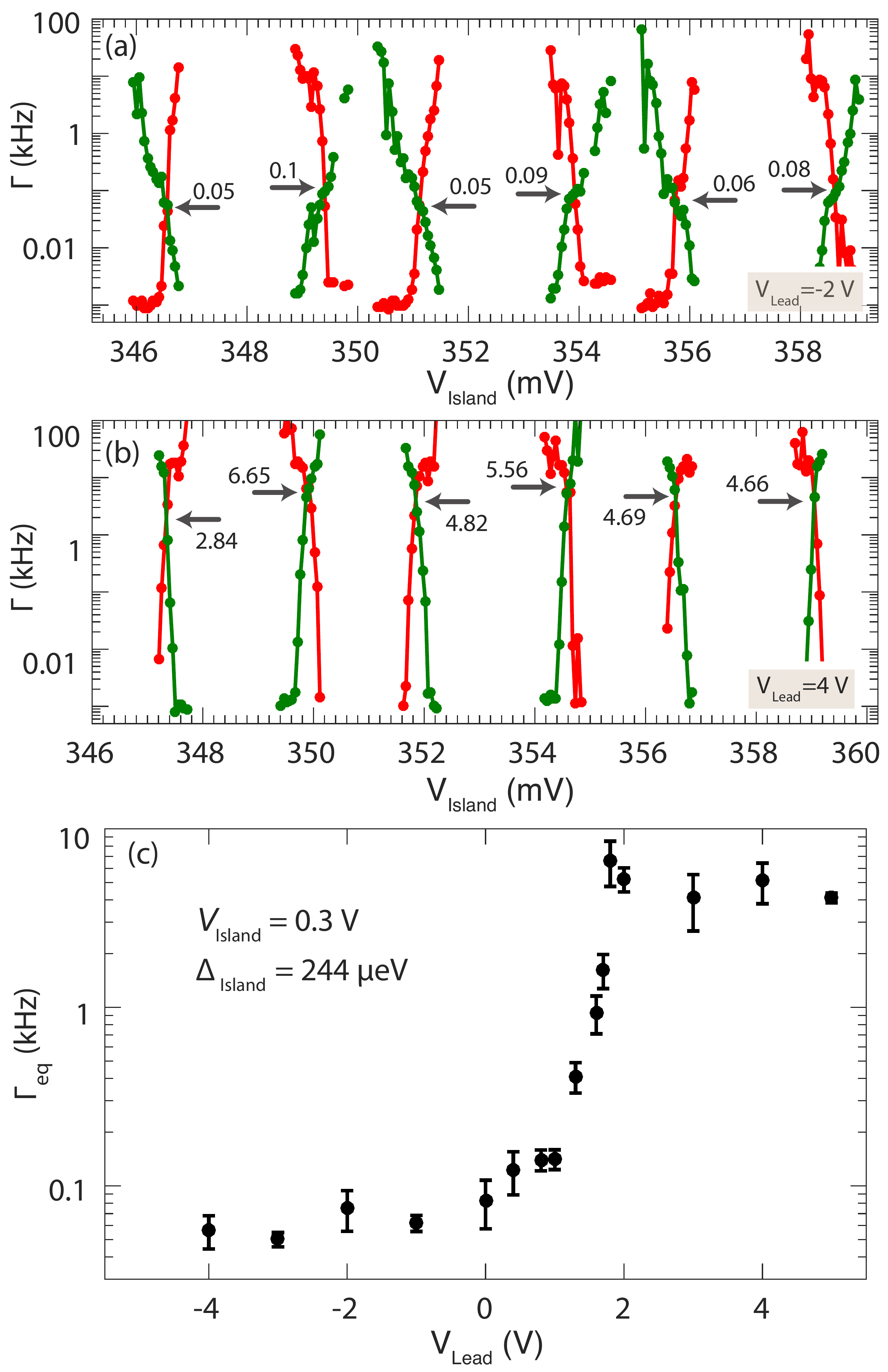}
\caption{(a,b) Tunneling rates $\Gamma_{\text{e} \rightarrow \text{o}}$ (green) and $\Gamma_{\text{o} \rightarrow \text{e}}$ (red)  for (a) $V_{\rm{Lead}}=-2$~V. $V_{\rm{MCut}}=-2.53$~V keeps normal-state conductance of middle cutter at $\sim 0.35\,e^2/h$. (b) $V_{\rm{Lead}}=+4$~V. $V_{\rm{MCut}}=-3.03$~V keeps normal-state conductance of middle cutter at $\sim 0.35\,e^2/h$ as in (a). (c) Average equilibrium (on resonance) tunneling rate, $\Gamma_{\rm{eq}}$, as a function of $V_{\rm{Lead}}$ for $V_{\rm{Island}} \sim 0.3$~V. All data at zero magnetic field.}
\label{Fig3}
\end{center}
\end{figure}

Motivated by Refs.~\cite{AumentadoPRL04, PekolaAPL09}, which demonstrated that QPs are attracted to small-gap regions, we investigate how QP tunneling of the island depends on plunger voltage, $V_{\text{Lead}}$, which can alter the induced gap of the nanowire lead. To do so, we associate the QP tunneling rate with the equilibrium (on resonance) tunneling rate, $\Gamma_{\rm{eq}}$, found by averaging $\Gamma_{\text{e(o)} \rightarrow \text{o(e)}}$ [green(red)] over several adjacent charge degeneracies [arrows in Figs.~\ref{Fig2}(d)] for fixed $V_{\text{Lead}}$. 
Values for $\Gamma_{\rm{eq}}$ are determined in a similar manner for various values of $V_{\text{Lead}}$. 

Figures~\ref{Fig3}(a,b) show tunneling rates $\Gamma_{\text{e(o)} \rightarrow \text{o(e)}}$ as $V_{\text{Island}}$ drives the island through several Coulomb valleys for widely different lead plunger voltages, $V_{\text{Lead}}=-2$~V and $+4$~V. Because of unavoidable capacitive coupling of $V_{\text{Lead}}$ to the tunnel barrier, it is necessary to adjust  $V_{\text{MCut}}$ in order to keep the effective tunnel barrier constant. Otherwise, changes in $\Gamma_{\rm{eq}}$ could simply reflect changes in the barrier transmission with changing $V_{\text{Lead}}$. To compensate this cross-coupling, $V_{\text{MCut}}$ is adjusted whenever $V_{\text{Lead}}$ is changed, such that the normal state conductance for this barrier remains at $G_N~=~0.35$~$e^2/h$ using a separate transport measurement. For instance, in Fig.~\ref{Fig3}(a) $V_{\text{MCut}}=-2.53$~V, and in Fig.~\ref{Fig3}(b) $V_{\text{MCut}}=-3.03$~V. The average tunneling rate $\Gamma_{\rm{eq}}$ at crossing points is $\Gamma_{\rm{eq}}=5.1 \pm 1.3$~kHz for $V_{\text{Lead}} =+4$~V, and $\Gamma_{\rm{eq}}=50 \pm 30$~Hz for $V_{\text{Lead}}=-2$~V. In other words, the {\it compensated} increase of $V_{\text{Lead}}$ by 6 V between (a) and (b) increases the tunneling rate by two orders of magnitude without changing the normal-state resistance. 

Figure~\ref{Fig3}(c) shows $\Gamma_{\rm{eq}}$ as $V_{\text{Lead}}$ is varied from $-5$~V to $+5$~V, with $V_{\text{Island}}$ fixed near $+0.3$~V and the middle cutter compensated using $V_{\text{MCut}}$ as described above. Resonances that depend on $V_{\text{MCut}}$,  presumably due to disorder in the middle-cutter junction, give rise to a non-monotonic dependence of tunneling with $V_{\text{MCut}}$ and corresponding non-monotonic normal state conductance $G_N$ as a function of $V_{\text{MCut}}$. Before the counting experiment, we open $V_{\text{LCut}}$ and $V_{\text{RCut}}$, and use transport measurement to verify that $G_N\sim 0.35\, e^2/h$ for each set of ($V_{\text{Lead}}$, $V_{\text{MCut}}$) \cite{supp}. Figure~\ref{Fig3}(c) shows a change in $\Gamma_{\rm{eq}}$ by two orders of magnitude when the difference between $V_{\text{Island}}$ and $V_{\text{Lead}}$ is about 1~V. For larger gate-voltage differences, $|V_{\text{Island}}-V_{\text{Lead}}|>1$ V, the tunneling rate saturates at a low and high value with little gate dependence. 

\begin{figure}[t]
\begin{center}
\includegraphics[width=3in,keepaspectratio]{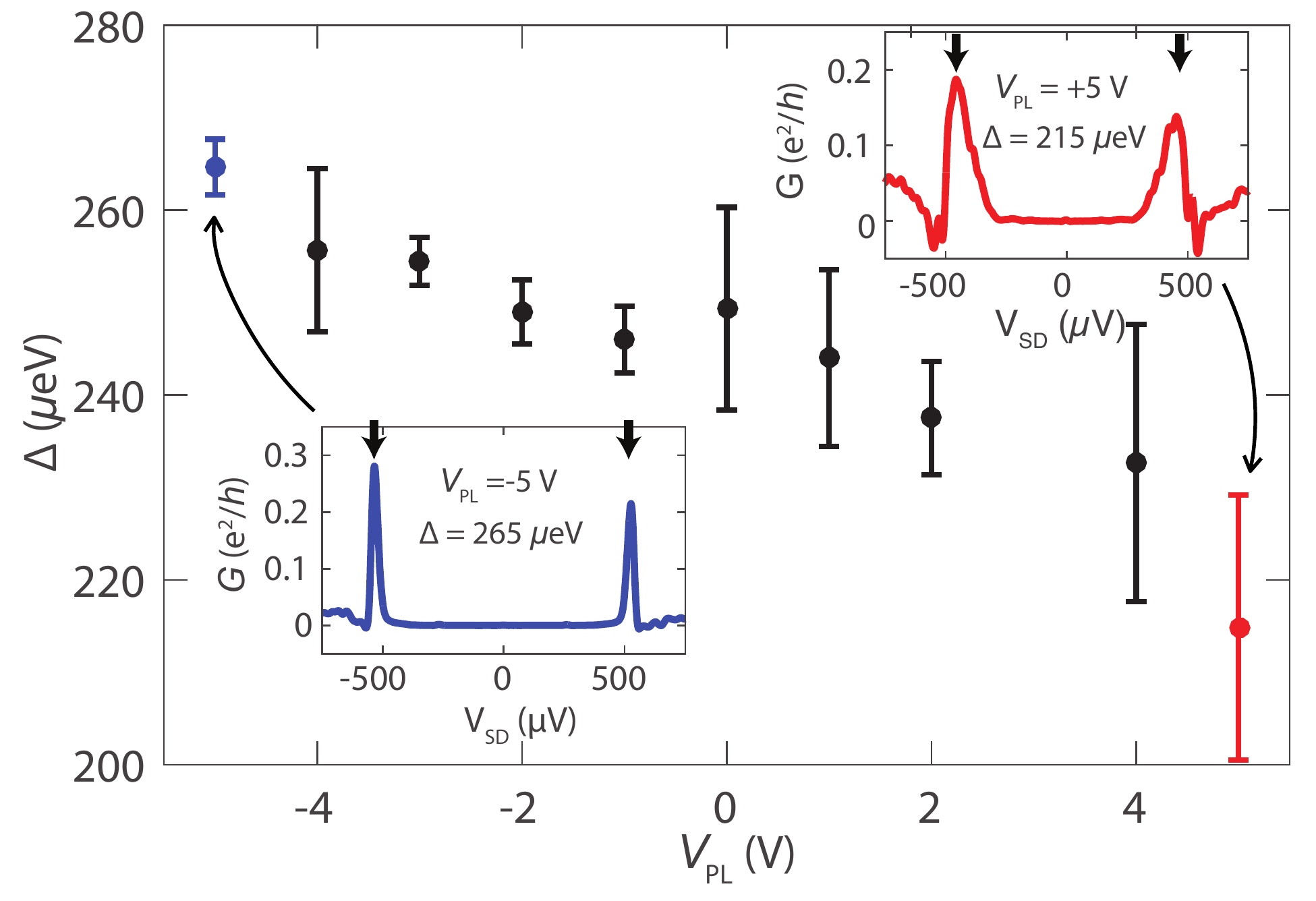}
\caption{Dependence of induced gap $\Delta$ on gate voltage $V_{\text{PL}}$ applied to all plunger gates (see text). $\Delta$ is measured by forming an SIS junction with the middle cutter while the other cutters are open by applying positive gate voltages. $\Delta$ is determined from the positions of the peaks in the SIS differential conductance, $G$, as a function of voltage bias, $V_\text{SD}$. Insets show two d$I$/d$V$ bias traces measured at $V_{\text{PL}} =\pm 5$~V. All data at zero magnetic field.}
\label{Fig4}
\end{center}
\end{figure}

\begin{figure}[h]
\begin{center}
\includegraphics[width=2.9in,keepaspectratio]{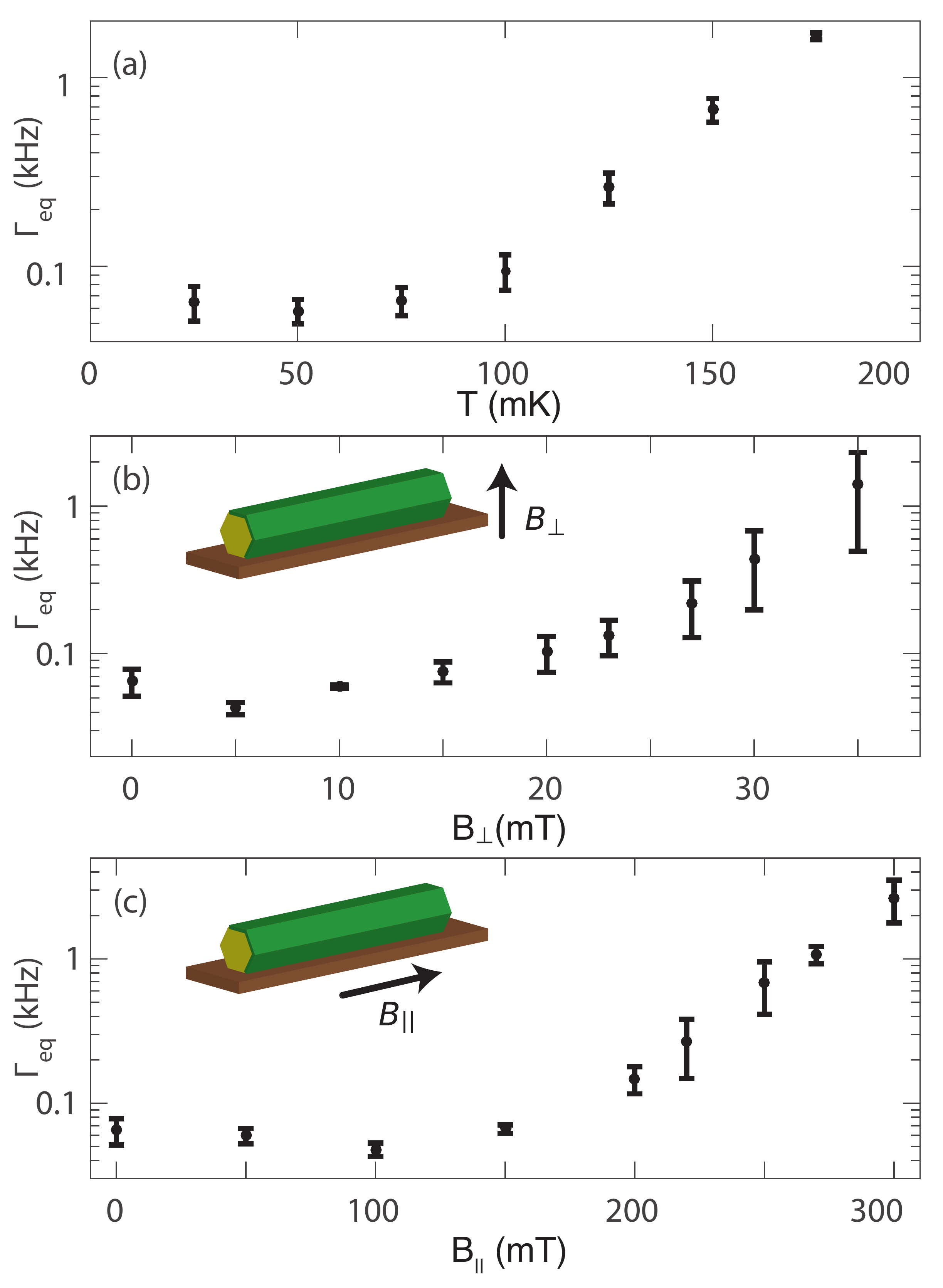}
\caption{Average tunneling rate $\Gamma_{\rm{eq}}$ as a function of (a) temperature at zero magnetic field, (b) magnetic field perpendicular to the substrate at base temperature, and (c) axial field along the nanowire at base temperature for the same gate configuration as in Fig.~\ref{Fig2}. Increasing the temperature leads to a reduction and softening of the induced gap, increasing the quasiparticle population. Similar interpretations can be made for the field dependences.}
\label{Fig5}
\end{center}
\end{figure}

Previous experiments on similar NWs have demonstrated that the induced superconducting gap in the NW can be controlled by the plunger gate voltage \cite{ChangNNano}. Our interpretation of the origin of the changes in $\Gamma_{\rm{eq}}$ with $V_{\text{Lead}}$ is that at more positive values of $V_{\text{Lead}}$ the superconducting gap is smaller on the lead than in the island, $\Delta_{\text{Lead}}<\Delta_{\text{Island}}$. In addition, as $V_{\text{Lead}}$ becomes more positive there is a softening of the induced superconducting gap. Both of these effects are evident in SIS transport measurements performed on the tunnel junction between the island and the lead. The distance between the two coherence peaks, marked by two black arrows in Fig.~\ref{Fig4} insets, gives $2(\Delta_\mathrm{Island}+\Delta_\mathrm{Lead})$, where $\Delta_{\mathrm{Island\,(Lead)}}$ is the superconducting gap on the island (lead) side of the tunnel barrier. We set all plungers $V_{\text{L}}$, $V_{\text{Island}}$, $V_{\text{Lead}}$, and $V_{\text{R}}$ to the same value, denoted $V_{\text{PL}}$, and consider $\Delta\sim\Delta_\mathrm{Island}\sim\Delta_\mathrm{Lead}$. Figure \ref{Fig4} shows the change in induced superconducting gap $\Delta$ while varying $V_{\text{PL}}$.  
The induced superconducting gap decreases linearly from 270 to 210 $\mu$eV as the plunger voltage is increased. 
This is because an increase in gate voltage increases the electron density in the semiconductor, which weakens the proximity effect induced from the ultrathin Al layer. 
Based on the data in Fig.~\ref{Fig4}, the drastic change in $\Gamma_{\rm{eq}}$ in Fig.~\ref{Fig3}(c) over a voltage range $|V_{\text{Island}}-V_{\text{Lead}}|\simeq1$ V corresponds to a change of superconducting gap of $\Delta_{\text{Island}}-\Delta_{\text{Lead}} \simeq 5$ $\mu$eV. 
This energy matches approximately the base temperature of our cryostat, suggesting that thermal smearing governs the cross-over from low to high tunneling rates in the observed $\Gamma_{\rm{eq}}(V_{\text{Lead}})$.

Single-charge tunneling rates were also investigated as a function of the temperature and magnetic field. Keeping the same gate configuration as in Fig.~\ref{Fig2}, we repeat the counting measurement. Inside a Coulomb valley, the charge state is always stable, similar to Fig.~\ref{Fig2}(a). Next, we focus on the tunneling rate at degeneracy as introduced above. $\Gamma_{\rm{eq}}$ increases with temperature [Fig.~\ref{Fig5}(a)] and magnetic field applied perpendicular to the substrate [Fig.~\ref{Fig5}(b)] and along the NW [Fig.~\ref{Fig5}(c)]. The increased rate as a function of these parameters is consistent with the softening and reduction of the induced superconducting gap leading to an enhancement of QP generation rates and $\Gamma_{\rm{eq}}$. At the highest measured axial magnetic field value in Fig.~\ref{Fig2}(b), $B_{||}$~=~300~mT, the island changes its ground state configuration approximately every 300~$\mu$s ($\Gamma_{\rm{eq}}\approx$~3~kHz), similar to metallic devices \cite{MaisiPRL14, DelsingPRB08,ClarkPRB08}. 

In summary, we investigated a gate-defined Coulomb island in the InAs/Al nanowire such that quasiparticles can only tunnel from one side of the island. Employing reflectometry, we count tunneling events on an island through that tunnel barrier in real time. Deep in the Coulomb valley the island shows no signal of quasiparticle tunneling on time scales ranging from sub-microsecond time scales to hours \cite{MannilaNPhys}. At charge degeneracy points, the tunneling rate varies by orders of magnitude with electrostatic gating of the island (Fig.~\ref{Fig3}c) as well as the lead (Supplemental Fig.~\ref{FigSupp1}). We interpret the dependence as arising from the gate dependence of the {\it relative} sizes of the induced gaps in the island and lead \cite{Mannila, AumentadoPRB06, FergusonPRL06} as well as the softness of the induced gaps. Tunneling rates also shows a strong dependence temperature and magnetic field, effects not yet modeled.

We thank Roman Lutchyn, Dmitry Pikulin, Judith Suter, and Jukka Vayrynen for valuable discussions, and Shiv Upadhyay for help with fabrication. Research is supported by Microsoft, the Danish National Research Foundation, and the European Research Commission, grant 716655, and a grant (Project 43951) from VILLUM FONDEN.

\clearpage
\onecolumngrid

\begin{center}
{\bf SUPPLEMENTARY MATERIAL}
\end{center}

\setcounter{equation}{0}
\setcounter{figure}{0}
\setcounter{table}{0}
\setcounter{page}{1}
\makeatletter
\renewcommand{\theequation}{S\arabic{equation}}
\renewcommand{\thefigure}{S\arabic{figure}}

\subsection{Equilibrium poisoning rate $\Gamma_{\rm{eq}}$ as a function of $V_{\text{MCut}}$}

\begin{figure}[h]
\begin{center}
\includegraphics[width=4in,keepaspectratio]{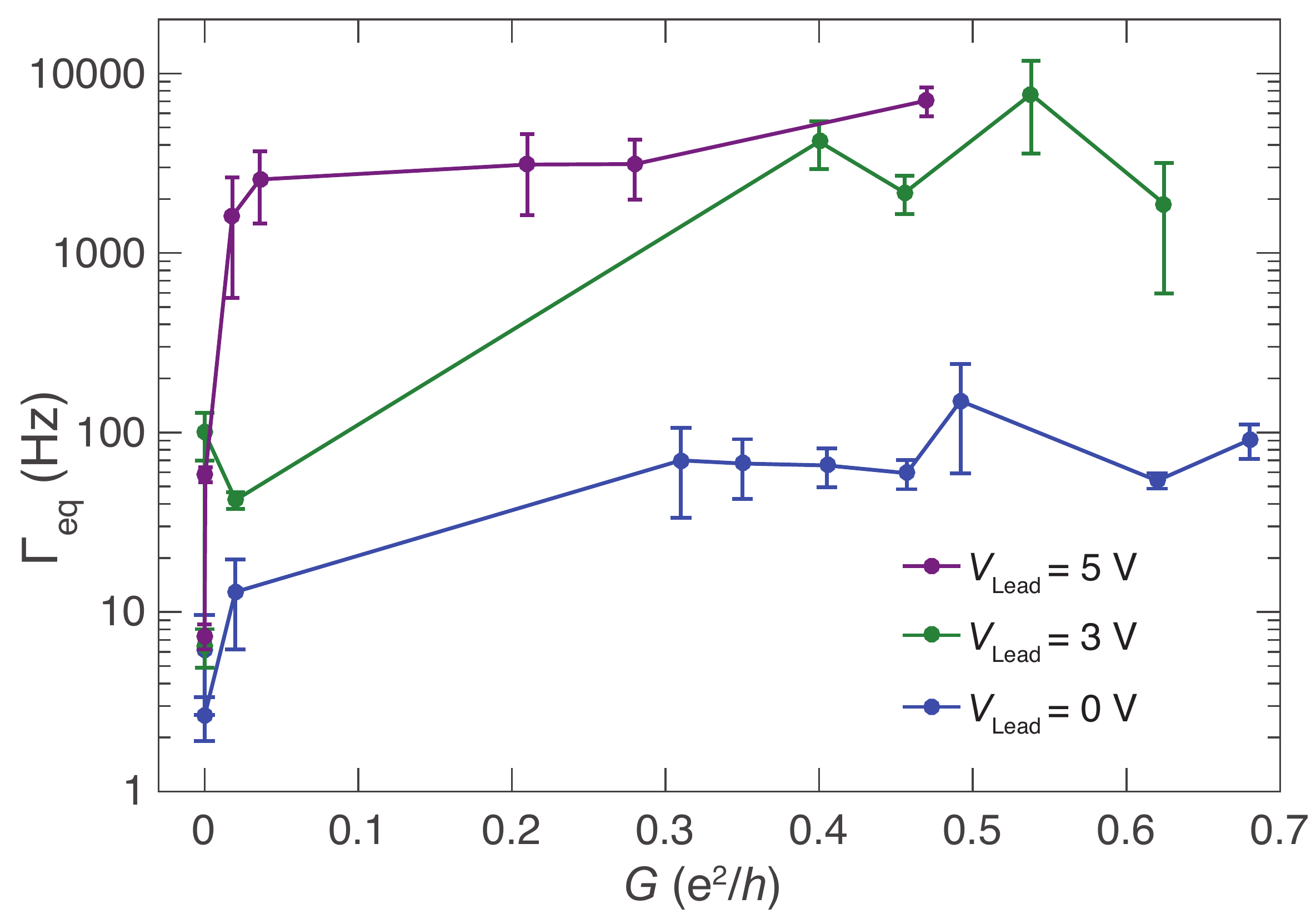}
\caption{Quasiparticle poisoning rate $\Gamma_{\rm{eq}}$ as a function of device conductance $G$ controlled by voltage $V_{\text{MCut}}$ for three different values of $V_{\text{Lead}}$. For these measurements $V_{\text{LCut}}$ is set to sufficiently negative values such that tunnelling to the left lead is negligible, while $V_{\text{Island}}$ is stepped over a range of approximately 300~mV, corresponding to a few Coulomb valleys. For a wide range of $G$ we observe that the poisoning rate can be controlled by $V_{\text{Lead}}$.}
\label{FigSupp1}
\end{center}
\end{figure}

We repeat the quasiparticle counting experiment at different voltages on the lead gate $V_{\text{Lead}}$ as a function of cutter voltage $V_{\text{MCut}}$. In Fig.~\ref{FigSupp1}, the poisoning rate $\Gamma_{\rm{eq}}$ is plotted as a function of high-bias conductance $G$ for three different lead voltages $V_{\text{Lead}}~=~0$~V, $V_{\text{Lead}}~=~3$~V, and $V_{\text{Lead}}~=~5$~V. When $V_{\text{MCut}}$ is set to negative voltages (closed), there is no electron tunneling into the island, as $V_{\text{LCut}}$ is also closed throughout the experiment. The time traces show a constant signal with no switching behavior. As $V_{\text{MCut}}$ is opened, the poisoning rate $\Gamma_{\rm{eq}}$ increases and quickly reaches a saturation plateau. The effect of disorder does not play a major role in the tunneling rate of electrons but introduces only local non-monotonicity in $\Gamma_{\rm{eq}}$  dependence on tunnel-barrier conductance $G$. The plateau in Fig.~3(c) in the main text holds for a large voltage range on the cutter $V_{\text{MCut}}$, and as a result allows the increase in the poisoning rate $\Gamma_{\rm{eq}}$ to be associated with the difference between $V_{\text{Lead}}$ and $V_{\text{Island}}$.

\subsection{Absence of quasiparticle poisoning as a function of temperature and fields}

\begin{figure}[h]
\begin{center}
\includegraphics[width=5in,keepaspectratio]{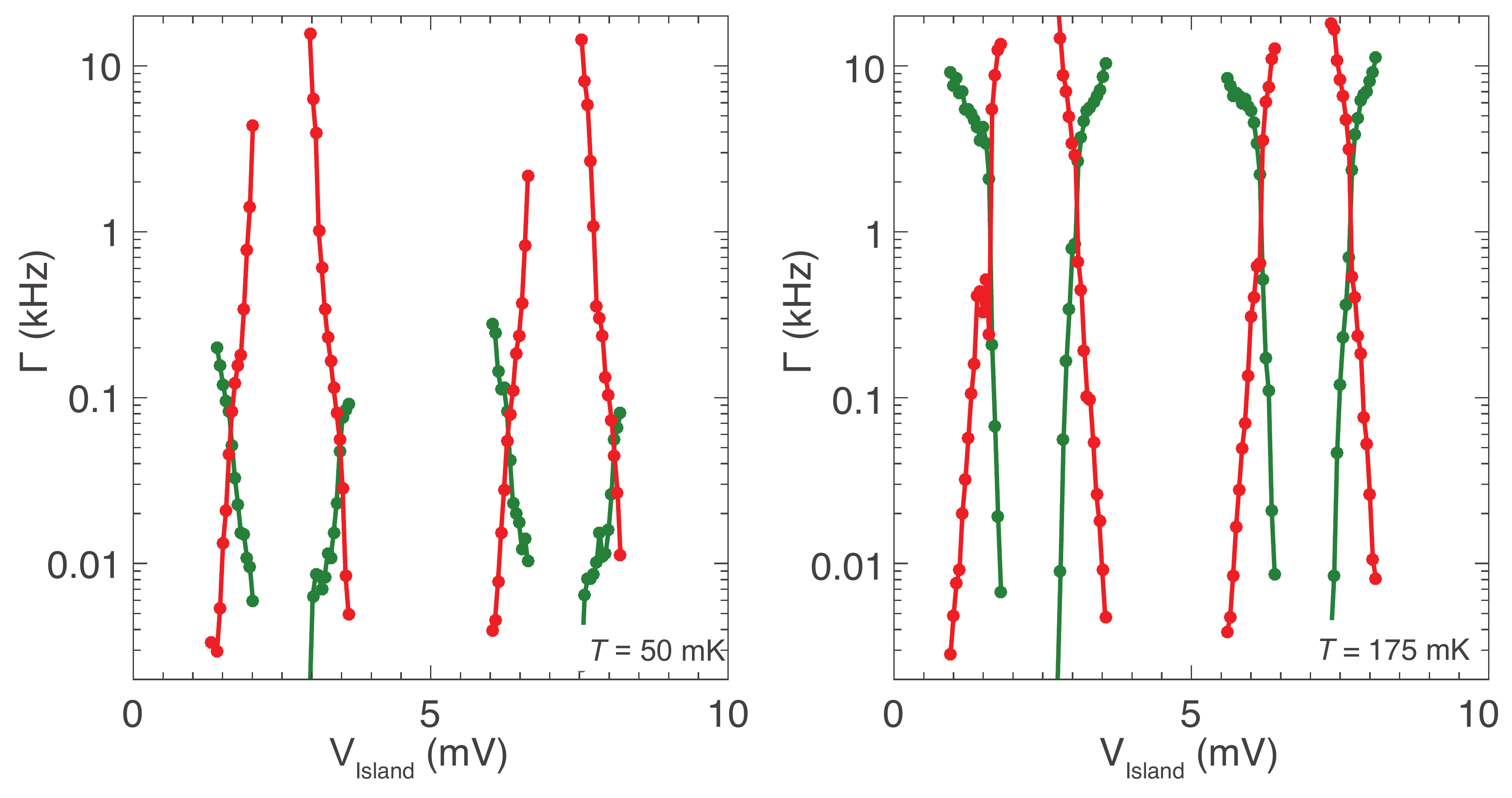}
\caption{Tunneling rates $\Gamma$ at zero magnetic field as a function of island plunger voltage $V_{\text{Island}}$ at a temperature of (a) 50 mK and (b) 175 mK. Tunneling rates vanish within Coulomb valleys and increase with temperature at charge degeneracies.}
\label{FigSupp2}
\end{center}
\end{figure}

\begin{figure}[h]
\begin{center}
\includegraphics[width=5in,keepaspectratio]{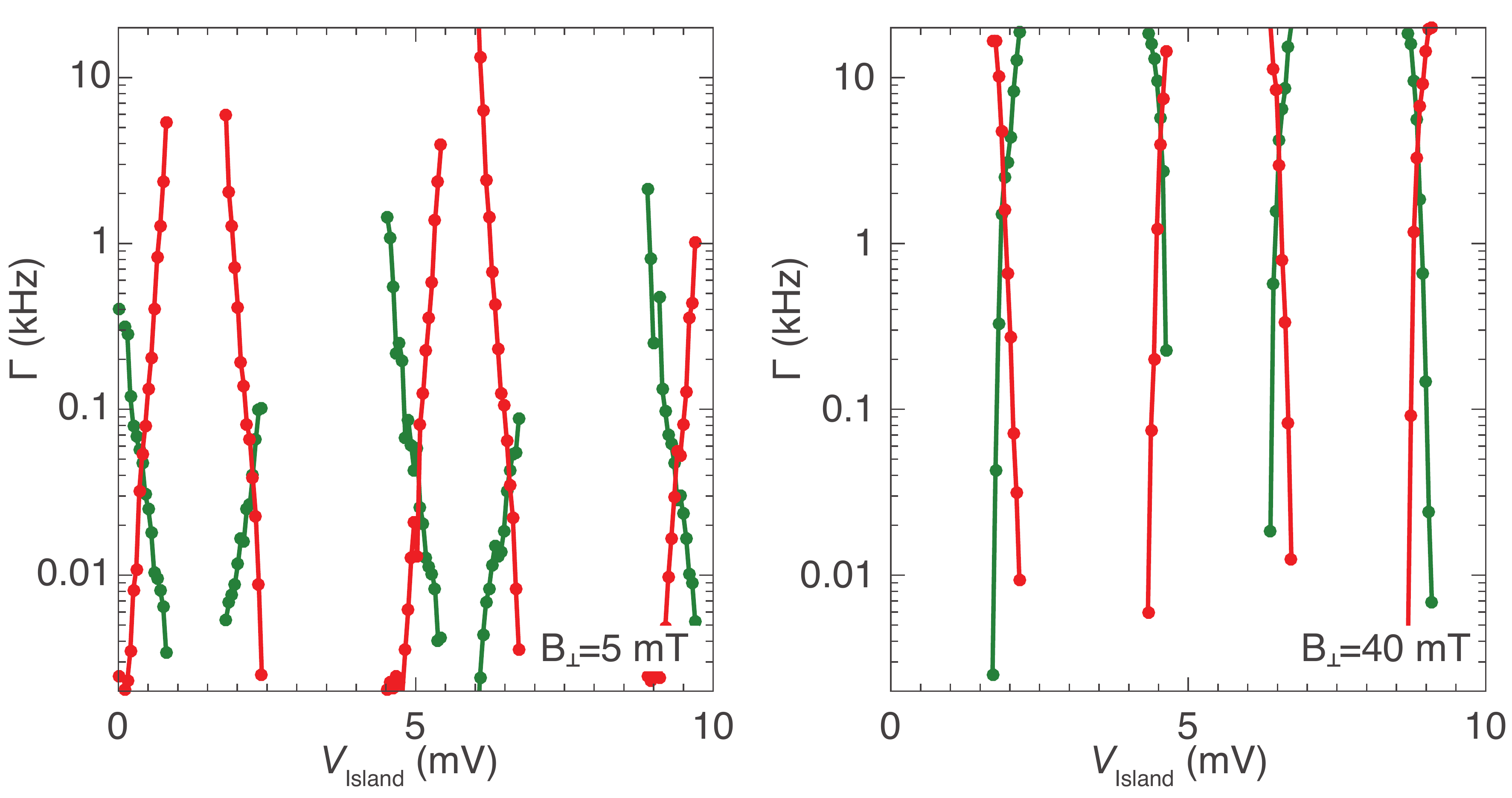}
\caption{Tunneling rates $\Gamma$ (measured at base temperature of 30 mK) as a function of island plunger voltage $V_{\text{Island}}$ in a perpendicular magnetic field of (a) 5 mT and (b) 40 mT. Tunneling rates vanish within the Coulomb valleys and increase at charge degeneracies with perpendicular field, consistent with the softening of the induced superconducting gap.}
\label{FigSupp3}
\end{center}
\end{figure}

\begin{figure}[h]
\begin{center}
\includegraphics[width=5in,keepaspectratio]{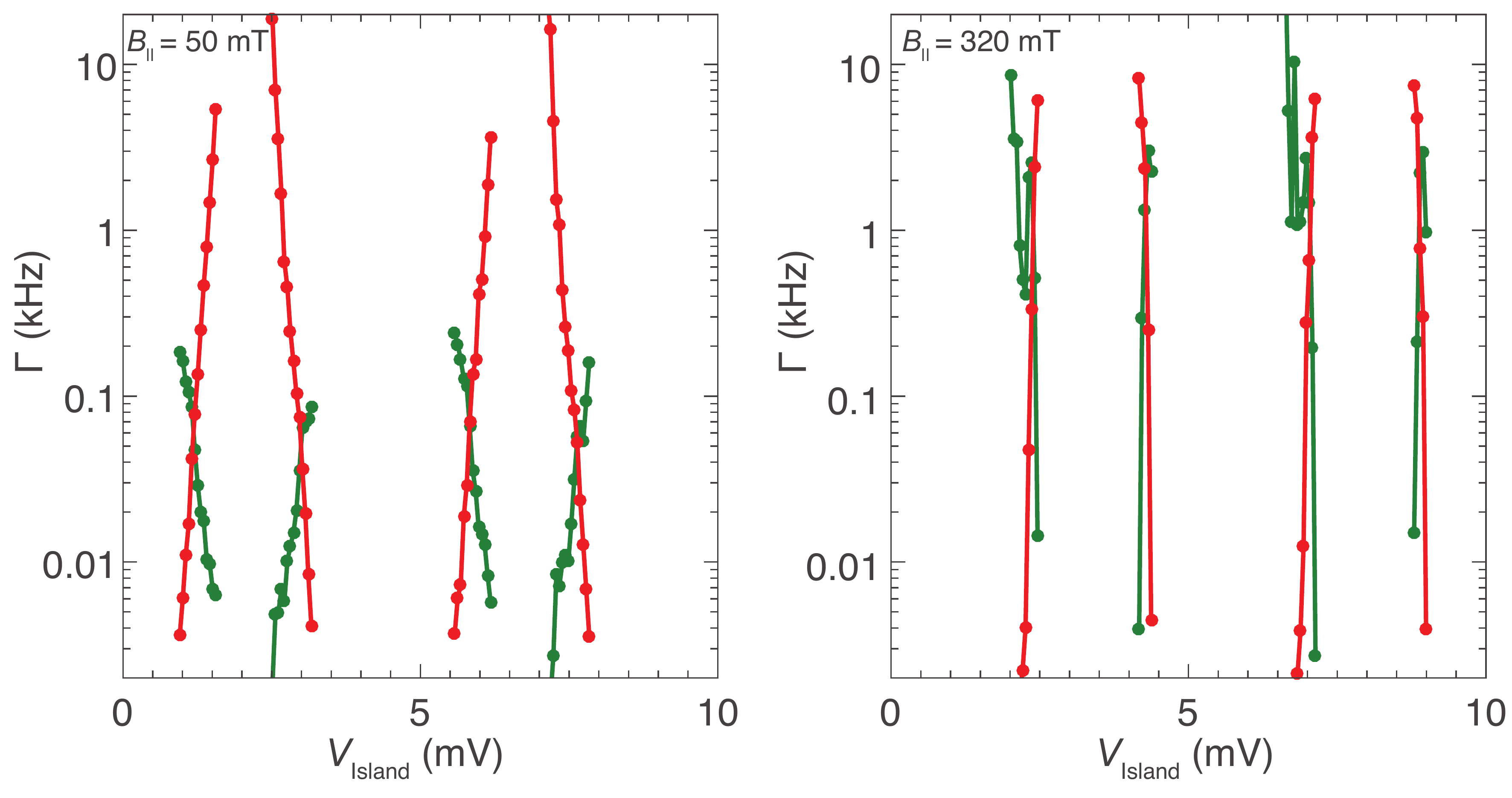}
\caption{Tunneling rates $\Gamma$ (measured at base temperature of 30 mK) as a function of island plunger voltage $V_{\text{Island}}$ in a parallel magnetic field of (a) 50~mT and (b) 320 mT. 
The even-odd effect is still visible at $B_{\parallel}$~=~320~mT.}
\label{FigSupp4}
\end{center}
\end{figure}

To emphasize the absence of quasiparticle poisoning in the hybrid InAs/Al island, we next present the study of tunneling rates as a function of the island plunger voltage $V_{\text{Island}}$ when the device is in a similar configuration to Fig.~2 presented in the main text. For this study the voltage $V_{\text{LCut}}$ is again set to a very negative value such that the tunnelling rate to the left lead is negligible, whereas $V_{\text{MCut}}$ is set such that the high-bias conductance through the junction is $G=0.35$ $e^2/h$  and $V_{\text{Lead}}$~=~0~V. In all three experiments where we change temperature (Fig.~\ref{FigSupp2}), perpendicular magnetic field (Fig.~\ref{FigSupp3}), and parallel magnetic field (Fig.~\ref{FigSupp4}), the Coulomb valleys always show small tunnelling rates indicating that the island rarely changes its parity when deep in the Coulomb valley. At the charge degeneracy points a rate increase is observed when increasing temperature or magnetic field, consistent with the softening of the superconducting gap. For all three dependencies shown below the measurements were stopped when the tunneling rates became too fast to measure.

\begin{thebibliography}{70}
\bibitem{SNR} The signal to noise ratio is defined as SNR = $\mu/\sqrt{2}\sigma$, with $\mu$ denotes the mean value and $\sigma$ denotes the standard deviation of the data.
\end{thebibliography}
\end{document}